\documentclass[traditabstract]{aa} % for the abstract without structuration 
\usepackage{graphicx}
\usepackage{txfonts}
\usepackage{natbib}
\begin{document}
\title{Enhanced [CII] emission in a $z$=4.76 submillimetre galaxy}

\author{Carlos De Breuck\inst{1} \and
Roberto Maiolino \inst{2} \and
Paola Caselli \inst{3} \and
Kristen Coppin \inst{4} \and
Steve Hailey-Dunsheath \inst{5} \and
Tohru Nagao \inst{6}}

\institute{European Southern Observatory, Karl Schwarzschild Stra\ss e 2, 85748 Garching, Germany \email{cdebreuc@eso.org} \and
INAF - Osservatorio Astronomico di Roma, via di Frascati 33, 00040 Monte Porzio Catone, Italy \and
School of Physics and Astronomy, University of Leeds, Leeds LS2 9JT, United Kingdom \and
Department of Physics, McGill University, Ernest Rutherford Building, 3600 Rue University, Montr\'{e}al, Qu\'{e}bec, H3A 2T8, Canada \and
Max-Planck-Institute for Extraterrestrial Physics, Giessenbachstra\ss e 1, 85748 Garching, Germany \and
Research Center for Space and Cosmic Evolution, Ehime University, 2-5 Bunkyo-cho, Matsuyama 790-8577, Japan
}

   \date{Received 2011 March 11; accepted 2011 April 27}

% \abstract{}{}{}{}{} 
% 5 {} token are mandatory
 
  \abstract
   {We present the detection of bright [CII] emission in the $z$=4.76 submillimetre galaxy LESS~J033229.4-275619 using the Atacama Pathfinder EXperiment. This represents the highest redshift [CII] detection in a submm selected, star-formation dominated system. The AGN contributions to the [CII] and far-infrared (FIR) luminosities are small. We find an atomic mass derived from [CII] comparable to the molecular mass derived from CO. The ratio of the [CII], CO and FIR luminosities imply a radiation field strength $G_0$$\sim$10$^3$ and a density $\sim$10$^4$\,cm$^{-3}$ in a kpc-scale starburst, as seen in local and high redshift starbursts. The high $L_{\rm [CII]}/L_{\rm FIR}$=2.4$\times$10$^{-3}$ and the very high $L_{\rm [CII]}/L_{\rm CO(1-0)}$$\approx$10$^4$ are reminiscent of low metallicity dwarf galaxies, suggesting that the highest redshift star-forming galaxies may also be characterised by lower metallicities. We discuss the implications of a reduced metallicity on studies of the gas reservoirs, and conclude that especially at very high redshift, [CII] may be a more powerful and reliable tracer of the interstellar matter than CO.}


   \maketitle
%
%________________________________________________________________

\section{Introduction}

%It is now well established that the bulk of stars in galaxies were formed at $z>2$ \citep[e.g.][]{madau96,lilly96}. 
Blank field submillimetre (submm) surveys have uncovered a population of distant galaxies undergoing a major epoch of star formation. To date, hundreds of these submm galaxies (SMGs) are known \citep[e.g.][]{coppin06,weiss09}. However, we only have basic knowledge of the star formation process in SMGs, and how it compares with star formation regions in the local universe. The most obvious tracer of the cold molecular gas from which stars form is CO, which has been extensively observed in SMGs \citep[e.g.][]{greve05,tacconi06}, and has now also been detected in other high redshift galaxies \citep[e.g.][]{daddi09,tacconi10}. These CO observations have shown that SMGs are gas rich systems ($M(H_2)=10^{10}-10^{11}$M$_{\odot}$), providing sufficient fuel to sustain massive star formation rates exceeding 1000\,M$_{\odot}$yr$^{-1}$. While the CO lines offers numerous advantages, notably their brightness allowing their emission to be spatially resolved \citep[e.g.][]{tacconi08}, there are also some complications. First, there are uncertainties of a factor five in the CO to H$_2$ conversion factor $X_{\rm CO}$, especially when integrated over an entire galaxy where significant variations in $X_{\rm CO}$ are expected \citep{shetty11}. Second, the derived gas masses depend on the excitation conditions, and a well sampled CO spectral line energy distribution is required to determine the full distribution of temperature and density components \citep[e.g.][]{weiss07,danielson11}.
Third, significant dust optical depth effects \citep{papadopoulos10a} may suppress the high excitation lines, while contributions from shock excitation \citep{papadopoulos08} may boost them.

The [CII]~$\lambda$157.74\,$\mu$m line has recently emerged as a powerful alternative line to study the interstellar medium in high redshift objects. It arises from photodissociation regions (PDRs) associated with star-forming regions and is the dominant cooling line, representing up to 1\% of the total luminosity \citep[e.g.][]{crawford85,stacey91}. The [CII] line is nearly always optically thin, though it may also be attenuated by dust extinction. [CII] detections have now been reported in five $z>2$ objects \citep{maiolino05,iono06,maiolino09,wagg10,ivison10}, and 12 galaxies at $1$$<$$z$$<$$2$ \citep{hailey10,stacey10}. The first high-redshift [CII] observations seemed to confirm the trend seen in local galaxies, where the ratio of [CII] and far-infrared (FIR) luminosity $L_{\rm [CII]}/L_{\rm FIR}$ is lower by about an order of magnitude for sources with $L_{\rm FIR}>10^{11}$L$_{\odot}$ \citep{luhman98,luhman03}. However, subsequent observations revealed that about half of high redshift [CII] lines have equal relative brightness than nearby normal galaxies. Recently, \citet{gracia11} showed that $z$$>$1 and local galaxies overlap when using $L_{\rm FIR}/M_{\rm H_2}$ ratio instead of $L_{\rm FIR}$ as a proxy for $L_{\rm [CII]}/L_{\rm FIR}$, probably because $L_{\rm FIR}/M_{\rm H_2}$ is more closely related to the properties of the clouds than $L_{\rm FIR}$ alone. 
%Using the redshift and Early Universe Spectrometer (ZEUS), 
The improved [CII] statistics also allowed \citet{stacey10} to conclude that starburst dominated systems have similar $L_{\rm [CII]}/L_{\rm FIR}$ than local normal galaxies, while AGN dominated systems have lower ratios like those in local ultra-luminous infrared galaxies (ULIRGs). Both classes imply kpc-size emitting regions, but the AGN-dominated sources appear to have 8 times more intense far-UV radiation fields, which \citet{stacey10} speculate results from an AGN-triggered population of very young stars. \citet{maiolino09} suggested that lower metallicities may explain the increased relative [CII] brightness in the highest redshift sources compared to similarly bright FIR sources in the local universe.

Clearly, the presence of an AGN complicates the determination of the physical conditions in the star-forming regions of their host galaxies, as the AGN may contribute to both the FIR continuum and [CII]. 
%Recent deep X-ray studies of SMGs indicate that in $\sim$85\% of SMGs, the AGN does not dominate the bolometric luminosity, with the faint X-ray emission originating from star-forming processes rather than an AGN \citep{georgantopoulos11,laird10}.  
%Moreover, the growth of the black hole appears to lag that of the host galaxy in SMGs \citep{alexander05,alexander08}. Any AGN activity in the most distant SMGs is therefore likely to have a smaller influence on their host galaxies.
AGNs may also enrich the ISM with metals \citep[e.g.][]{juarez09}, which would complicate studying the impact of metallicity in SMGs. In a recent {\it Herschel} study, \citet{santini10} found that 1$<$$z$$<$2 SMGs have higher dust masses than local ULIRGs, while their derived gas metallicities are lower. It is as yet unclear if this apparent discrepancy continues to the highest redshift SMGs, who are expected to have lower metallicities. In this letter, we present [CII] observations of LESS~J033229.4-275619 at $z$=4.755 \citep[hereinafter LESS~J033229.4;][]{coppin09}. After AzTEC-3 \citep[$z$=5.298;][]{riechers10,capak11}, this is the most distant SMG currently known. We detect bright [CII] emission and show it to be consistent with an extended starburst region, and having close characteristics to local sub-solar metallicity regions. Throughout, we assume $H_0=71~{\rm km~s}^{-1}~{\rm Mpc}^{-1}$, $\Omega _{\Lambda}=0.73$, and $\Omega _{\rm m}=0.27$. We define the far-IR luminosity as integrated over 42.5-122.5\,$\mu$m, to allow an easy comparison with \citet{hailey10} and \citet{stacey10}.

%__________________________________________________________________

\section{Observations and Results}
LESS~J033229.4 was observed with the Swedish Heterodyne Facility Instrument \citep[SHFI;][]{vassilev08} on the Atacama Pathfinder EXperiment \citep[APEX\footnote{APEX is a collaboration between the Max-Planck-Institut fur Radioastronomie, the European Southern Observatory, and the Onsala Space Observatory.};][]{guesten06} 12\,m telescope in six observing runs, from 2010 April 16 to August 24 (run ID 085.A-0584(A)), for a total of 14.5\,h of observations, including sky observations, calibrations and overheads. The weather conditions were generally very good with precipitable water vapour 0.4$<$PWV$<$1.0\,mm. We used the APEX-2 receiver tuned to 330.242937\,GHz (in lower sideband), which is the expected frequency of the [CII]$\lambda$157.741\,$\mu $m line at the redshift z=4.762 provided by the CO(2-1) detection \citep{coppin10}. The two units of the Fast Fourier Transfer Spectrometer (FFTS) backend were offset by $\pm$200\,km\,s$^{-1}$ to increase the velocity coverage. However, as the line was well centred at the expected position, we discarded the outer 200\,km\,s$^{-1}$ covered only by a single FFTS as we found them to be noisier and dominated by baseline subtraction effects. Observations were done in wobbler-switching mode, with a symmetrical azimuthal throw of 20\arcsec\ and a frequency of 0.5\,Hz. Pointing was checked on the nearby source R-For every $\sim$1\,h and found to be better than 3\arcsec\ (with a beam size of 20\arcsec). The focus was checked on the available planets, especially after sunrise when the telescope deformations are largest.

The data were analyzed by using CLASS (within the GILDAS-IRAM package), following the procedures described by \citet{maiolino09}. In short, we determined the noise in the individual spectra using a zero order baseline, and removed the noisiest 20\%. We subtracted a first order baseline to the combined spectrum by interpolating the channels in the line-free velocity ranges $-$850$<$$\Delta V$$<$$-$600\,km\,s$^{-1}$ and $-$250$<$$\Delta V$$<$$-$50\,km\,s$^{-1}$. The final spectrum includes a total of 2.5\,h on-source integration time.

% Figure 1
\begin{figure}
\centering
\includegraphics[width=9cm]{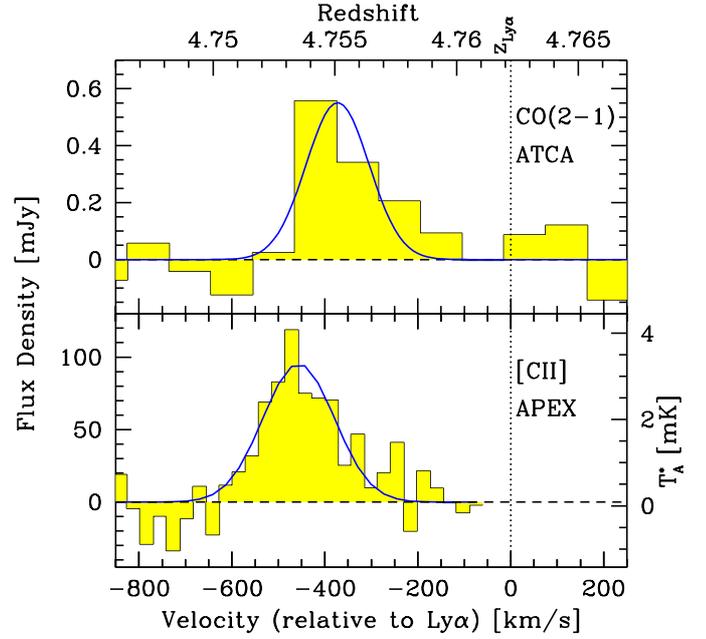}
\caption{{\it Top panel:} ATCA CO(2-1) spectrum of LESS~J033229.4 binned to 90\,km\,s$^{-1}$ \citep{coppin10}. {\it Bottom panel:} [CII] spectrum obtained with SHFI/APEX binned to 28\,km\,s$^{-1}$ (31\,MHz). The blue solid lines shows the best fitting Gaussian profiles. The bottom axis shows the velocity offset relative to the Ly$\alpha$ redshift of \citet{coppin09}, which is also marked as a dotted vertical line. }
\label{co_cii}
\end{figure}

Figure~\ref{co_cii} shows the resulting spectrum. The [CII] line is detected at the 7.6$\sigma$ level. The significance of our detection is further strengthened by the correspondence in line width with the CO(2-1) detection. Table~\ref{properties} compares the redshifts of the Ly$\alpha$, [OII], CO(2-1) and [CII] lines, derived from single Gaussian fits. The [CII] is offset by $-$88$\pm$74\,km\,s$^{-1}$, $+$130$\pm$280\,km\,s$^{-1}$ and $-$470$\pm$120\,km\,s$^{-1}$ from CO(2-1), [OII] and Ly$\alpha$, respectively. The red velocity offset of the Ly$\alpha$ line has been observed in many high redshift objects \citep[e.g.][]{steidel10}, and can be explained by HI absorption in the interstellar medium surrounding the galaxy. The other 3 lines are more likely to trace the systemic redshift. Because (i) the velocity offset between [CII] and CO is only 1.2$\sigma$, and (ii) none of the $z$$>$1 [CII] detections show significant offsets, we ascribe the offset in LESS~J033229.4 to the poor S/N of the CO data rather than to a real physical velocity offset between the [CII] and CO emitting regions.

% Table 1
\begin{table}
\caption[]{Observed properties of LESS~J033229.4.}
\label{properties}
\begin{tabular}{lll}
Parameter & Value & Reference \\
\hline
$z_{\rm Ly\alpha}$ & 4.762$\pm$0.002 & \citet{coppin09} \\
$z_{\rm [OII]}$ & 4.751$\pm$0.005 & Alaghband-Zadeh et al., in prep. \\
$z_{\rm CO}$ & 4.755$\pm$0.001 & \citet{coppin10} \\
$z_{\rm [CII]}$ & 4.7534$\pm$0.0009 & this paper \\
$\Delta V_{\rm CO}$ & 160$\pm$65\,km\,s$^{-1}$ & \citet{coppin10} \\
$\Delta V_{\rm [CII]}$ & 161$\pm$45\,km\,s$^{-1}$ & this paper \\
$I_{\rm CO}$ & 0.09$\pm$0.02\,Jy\,km\,s$^{-1}$ & \citet{coppin10}\\
$I_{\rm [CII]}$ & 14.7$\pm$2.2\,Jy\,km\,s$^{-1}$ & this paper\\
$L_{\rm [CII]}$ & $1.02\pm0.15\times10^{10}$L$_{\odot}$ & this paper \\
$L_{\rm CO(2-1)}$ & $7.6\pm1.7\times10^{6}$L$_{\odot}$ & \citet{coppin10} \\
$L_{\rm FIR}^{\dagger}$ & $4.2\pm1.8 \times10^{12}$L$_{\odot}$ & \citet{coppin09} \\
\hline
\end{tabular}
$^{\dagger}$Converted using $L_{42.5-122.5\mu m}=L_{8-1000\mu m} / 1.45$ \citep{stacey10}.
\end{table}

\section{Discussion}

The [CII] emission from LESS~J033229.4 is among the brightest observed at $z$$>$2, in particular when compared to the CO emission. The observed $L_{\rm [CII]}/L_{\rm FIR}$=2.4$\times$10$^{-3}$ is similar to local normal galaxies rather than local ULIRGs. Even when considering\footnote{We use the $M(\rm H_2)$=1.6$\times$10$^{10}$\,M$_{\odot}$ from \citet{coppin10}, derived using $X_{\rm CO}$=0.8\,$\rm M_{\odot}\,(K\,km\,s^{-1}\,pc^2)^{-1}$.} $L_{\rm FIR}/M_{\rm H_2}$=260, LESS~J033229.4 still falls well above the relation defined by local and high redshift objects \citep[Figure~1 of][]{gracia11}. We now use the CO, [CII] and FIR luminosities to examine the physical conditions of the ISM in LESS~J033229.4 and try to explain the bright [CII] emission relative to other populations at both low and high redshift.

\begin{figure}
\centering
\includegraphics[width=9.2cm]{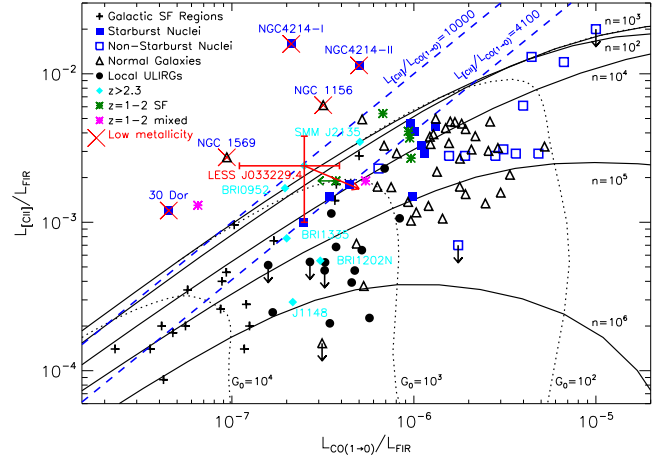}
\caption{$L_{\rm [CII]}/L_{\rm FIR}$ vs. $L_{\rm CO(1-0)}/L_{\rm FIR}$ for LESS~J033229.4 (red point with error bars dominated by the uncertainty in $L_{\rm FIR}$) compared with Galactic star-forming regions (crosses), starburst nuclei (filled squares), non-starburst nuclei (open squares), normal galaxies (triangles), local ULIRGs (circles), and high-redshift sources (asterisks and diamonds). Objects known to have low metallicity are overplotted by red crosses. Black lines represent the solar metallicity PDR model calculations for gas density ($n$) and FUV field strength ($G_0$) from \citet{kaufman99}. A vector indicates how galaxy data points are shifted for the PDR analysis in \S\,3. This figure is adapted from \citet{stacey10}, with additional data on BRI~1335-0417 ($z$=4.41) from \citet{riechers08} and \citet{wagg10}, and on the low metallicity galaxy NGC~4142 \citep{cormier10}. LESS~J033229 is located near the high $L_{\rm [CII]}/L_{\rm CO(1-0)}$ region occupied only by low metallicity galaxies.}
\label{cofir_ciifir}
\end{figure}

The $L_{\rm [CII]}/L_{\rm FIR}$ versus $L_{\rm CO(1-0)}/L_{\rm FIR}$ diagram is a particularly powerful tool as the two ratios are sensitive to gas density $n$ and the incident far-ultraviolet (FUV; 6\,eV$<$$h\nu$$<$13.6\,eV) flux $G_0$ \citep{stacey91,hailey10}. Figure~\ref{cofir_ciifir} compares LESS~J033229.4 with other objects and the solar metallicity PDR model curves of \citet{kaufman99}. This diagram can be used to roughly distinguish both the $n$ and $G_0$ appropriate for a source, but with some important caveats. 
First, the CO luminosity is expressed in terms of the ground-state transition of CO, while for LESS~J033229.4 and BRI~1335$-$0417\footnote{The [CII] detection of this source was published after \citet{stacey10}; we add the data of this target to Figure~\ref{cofir_ciifir}.}, only the (2-1) transition has been observed. For consistency with \citet{stacey10}, we assume $L_{\rm CO(2-1)}/L_{\rm CO(1-0)}$=7.2 (90\% of the thermalised, optically thick value) to calculate the $L_{\rm CO(1-0)}$. 
Second, Figure~\ref{cofir_ciifir} assumes that the contributions to the [CII] flux from the diffuse ionised medium and from cosmic rays (CR) heated gas are small compared to the ones from PDRs. Using PDR models, \citet{meijerink11} predict that the C$^0$ and C$^+$ abundances increase rapidly with the CR rates, especially when compared to CO. If the CR rate is related to the star formation rate \citep{papadopoulos10b}, this should thus be reflected in a higher [CII]/CO ratio, while powerful starbursts have similar ratios than normal galaxies (see Fig.~\ref{cofir_ciifir}). Any contribution from CR ionisation therefore does not appear to dominate the [CII]/CO ratio. Third, we assume that the AGN does not contribute significantly to the FIR and [CII] luminosity.  LESS~J033229.4 does contain a moderate luminosity AGN, revealed by (i) ultra-deep X-ray spectroscopy revealing a Compton-thick AGN \citep{gilli11}, and (ii) the detection of [NV]~$\lambda$1240\AA\ emission \citep{coppin09}. From their SED fitting, \citet{gilli11} derive an AGN bolometric luminosity of 7$\times 10^{45}$\,erg\,s$^{-1}$, which represents between 20 and 30\% of the total luminosity derived between 8 and 1000\,$\mu$m (and probably lower if the real $L_{\rm FIR}$ has been underestimated). It is therefore safe to assume that the FIR emission in LESS~J033229.4 is starburst dominated. The [CII] emission could also contain a contribution from the AGN through the X-ray Dominated Regions (XDRs) in the molecular torus surrounding the AGN. \citet{stacey10} estimate the XDR contributions using a scaling relation $L_{\rm [CII],AGN}$=2$\times$$10^{-3} L_{\rm X-ray}$. Using the intrisic (de-absorbed) rest-frame 2-10\,keV luminosity 2.5$\times 10^{44}$\,erg\,s$^{-1}$ \citep{gilli11}, we find that the predicted XDR contribution is $\sim$1.3\% of the observed [CII] lumonisity.
%The rest-frame UV spectral properties of LESS~J033229.4 resemble those of SMM~J02399$-$0136, but even in this object where the AGN is dominating in the UV, it appears not to dominate the FIR emission \citep{vernet01}. 

% Figure 2

In our PDR model analysis\footnote{We used the PDR toolbox code at http://dustem.astro.umd.edu/}, we multiply the CO flux by a factor of two to correct for geometry effects, and scale the [CII] flux by a factor 0.7 to remove non-PDR contributions \citep[shown by a vector in Fig.\,\ref{cofir_ciifir};][]{hailey10}. The geometry correction applies to all external galaxies where the FUV illumination comes from all sides, so this will not impact the {\it relative} position of LESS~J033229.4 compared to other galaxies.  We derive $n$$\approx$$10^4$\,cm$^{-3}$ and $G_0$$\approx$$10^3$. Such a FUV field is on the high end of those seen in low redshift normal galaxies, and both nearby and 1$<$$z$$<$2 starforming galaxies. However, it is significantly less intense than observed in Galactic OB star forming regions, three $z$$>$4 AGNs, and local ULIRGs \citep[though the latter may also be influenced by dust-bounded HII regions;][]{abel09,gracia11}. The uncertainties on the [CII]/FIR and CO/FIR ratios are large, but they are dominated by the uncertainty in $L_{\rm FIR}$, which may be 1.5 times higher \citep{coppin09}. Such an increase in $L_{\rm FIR}$ would shift the LESS~J033229.4 point to the lower left, along the line of constant $L_{\rm [CII]}/L_{\rm CO(1-0)}$=10$^4$. Similarly, an (unlikely) higher AGN contribution to $L_{\rm FIR}$ would shift the point to the top right. Neither shift would change $n$ or $G_0$ to a level that would change our conclusions.  Our results are thus broadly consistent with the findings of \citet{stacey10} who concluded that AGN dominated systems have $\sim$8 times higher FUV radiation fields than starburst dominated systems. The $n \approx 10^4$\,cm$^{-3}$ and $G_0 \approx 10^3$ imply a PDR surface temperature of $\sim$300\,K \citep{kaufman99}. Using formula~1 of \citet{hailey10}, we estimate the atomic mass associated with the C$^+$ region $M_a$=1.0$\pm$0.3$\times\,10^{10}$\,M$_{\odot}$, i.e. similar to the $M(\rm H_2)$=1.6$\pm$0.3$\times\,10^{10}$\,M$_{\odot}$ \citep{coppin10}.

The very high $L_{\rm [CII]}/L_{\rm CO(1-0)}$$\approx$10$^4$ places LESS~J033229.4 near the upper envelope of the solar metallicity plane-parallel slab PDR models in Figure~\ref{cofir_ciifir}. Only spherical PDR models can explain $L_{\rm [CII]}/L_{\rm CO(1-0)}$$>$$10^4$ ratios, in particular when the metallicity in reduced \citep[e.g.][]{bolatto99,rollig06}. Observational evidence is also building up that the high [CII]/CO region contains mainly low metallicity systems, where the [CII] is enhanced relative to CO and the FIR dust continuum. Extreme $L_{\rm [CII]}/L_{\rm CO(1-0)}$ ratios up to 7$\times$10$^4$ have only been observed in low metallicity dwarf galaxies \citep[e.g.][]{madden00}, and recently in star-forming regions of the low metallicity irregular galaxy NGC~4214 \citep[see Fig.~\ref{cofir_ciifir};][]{cormier10}. 
These high [CII]/CO ratios have been explained by the reduced dust abundance in low metallicity environments, which increases the mean free path of the FUV photons. This will not only decrease the FIR intensity and size of the CO cores, but also increase the size of the [CII] emitting regions as the photodissociating photons traverse a larger volume of the molecular cloud. This extended [CII] region may also contain self-shielded H$_2$, which can boost the CO-to-H$_2$ conversion factor by up to 100 times as seen in IC10 which has $\sim$1/5 solar metallicity \citep{madden97}. The total gas mass based only on CO observations may thus be substantially underestimated in these low metallicity systems.
%\citep[as indeed observed in the LMC and SMC;][]{israel11}. 
In contrast, the atomic gas mass derived from the [CII] line only scales linearly with the abundance, and is less affected.
While the $L_{\rm [CII]}/L_{\rm CO(1-0)}$ ratio in LESS~J033229.4 is not as extreme as in dwarf galaxies, such high ratios are still only observed in low metallicity environments. We therefore propose low metallicity as a natural explanation also in LESS~J033229.4. It is also remarkable that three of the six $z$$>$2 [CII] detections have similarly high $L_{\rm [CII]}/L_{\rm CO(1-0)}$ (Fig.~\ref{cofir_ciifir}), including BRI~0952$-$0115\footnote{The lensing amplification in BRI~0952$-$0115 also implies a less luminous AGN, which may explain why BRI~0952$-$0115 is closer to the starburst rather than AGN dominated systems in Fig.~\ref{cofir_ciifir}.} where \citet{maiolino09} already proposed low metallicity as an explanation for the large $L_{\rm [CII]}/L_{\rm FIR}$ ratio. Unfortunately, we do not have independent measurements of the global metallicity in LESS~J033229.4 or any of the high-$z$ objects with [CII] detections, though lower gas metallicities have indeed been observed in $z$$>$3 star forming galaxies \citep{maiolino08}. If, as expected, the highest redshift galaxies indeed have lower metallicities, their longer FUV mean free path lengths imply that for a constant FUV flux and density, the $G_0$ will be lower due to an increased geometrical dilution of the FIR flux \citep{israel11}. The [CII]/FIR ratio will thus also increase (Fig.~\ref{cofir_ciifir}). We conclude that [CII] rather than CO should be the prime tracer of the ISM in the highest redshift (non AGN-dominated) galaxies, because [CII] will not only be brighter, but also provides a more reliable determination of the total gas mass. 

The high redshift star-formation dominated objects have similar $G_0$ fields than local star forming regions (Fig.~\ref{cofir_ciifir}), but much higher FIR luminosities. So the former appear to be scaled up versions of local starbursts. As the FUV emission is absorbed by the dust in the molecular clouds, and re-radiated in the FIR, the extent of the emitting region for a given $L_{\rm FIR}$ will increase for smaller $G_0$. \citet{stacey10} use two scaling laws from \citet{wolfire90} to constrain this size: $G_0 \propto \lambda L_{\rm IR}/D^3$ if the mean free path of a FUV photon $\lambda$ is very small and $G_0 \propto L_{\rm IR}/D^2$ if the mean free path is very large. Applied to LESS~J033229.4, and assuming $G_0$$\sim$10$^3$, this yields a physical size $D_{\rm PDR}$=1.6--3.7\,kpc. As argued above, in low metallicity environments, the larger FUV mean free path approximation would be more appropriate, so the actual value is probably at the higher end of this range. This scale size is remarkably similar to the 4\,kpc full width half maximum of the stellar population derived from $K-$band imaging \citep{coppin10}, and shows that the starburst in LESS~J033229.4 is affecting the entire galaxy. Such kpc-scale starburst have been found by several authors \citep[e.g.][]{biggs08,carilli10}, with individual components as small as 100\,pc \citep{swinbank10}, and sometimes part of a merging system separated by a few kpc \citep{tacconi08}. However, the presence of a compact AGN, and differential lensing amplification of these different regions has complicated the interpretation. High spatial resolution [CII] and FIR observations of unlensed starburst dominated systems with the Atacama Large Millimeter and submillimeter Array will provide a major progress in this field. 

\section{Conclusions}
We have detected bright [CII] emission in a $z$=4.76 SMG, where the AGN contributes little the FIR and [CII] luminosities. We find an atomic mass derived from [CII] comparable to the molecular mass derived from CO. By similarity to local dwarf galaxies, the high $L_{\rm [CII]}/L_{\rm CO(1-0)}$$\approx$10$^4$ ratio suggests a low metallicity in the highest redshift star-forming galaxies. The enhanced [CII]  suggests that, as long as the AGN contribution is small, the [CII] line may be the most powerful and reliable tracer of the ISM.

\begin{acknowledgements}
We thank the anonymous referee for useful comments which have improved the paper, and the APEX/ESO staff for their expert help with the observations. KC acknowledges support from the the endowment of the Lorne Trottier Chair in Astrophysics and Cosmology at McGill, the National Science and Engineering Research Council of Canada, and the Centre of Research in Astrophysics of Qu\'{e}bec for a fellowship.
\end{acknowledgements}

\end{document}